\documentclass[a4paper]{article}

\usepackage[margin=1in]{geometry}

\usepackage[T1]{fontenc}
\newcommand{\changefont}[3]{
\fontfamily{#1} \fontseries{#2} \fontshape{#3} \selectfont}

\changefont{ptm}{m}{n}

\usepackage{setspace} 
\usepackage{graphicx}    

\usepackage{amsfonts,amssymb,amsmath}
\usepackage{graphics}
\usepackage{mathrsfs}
\usepackage{color}
\newtheorem{remark}{Remark}[section]

\newtheorem{theorem}{Theorem}[section]

\newtheorem{lemma}{Lemma}[section]

\long\def\symbolfootnote[#1]#2{\begingroup%
\def\thefootnote{\fnsymbol{footnote}}\footnote[#1]{#2}\endgroup} 

\begin{document}

\begin{center}
\Large \textbf{Persistence of Li-Yorke chaos in systems with relay}
\end{center}

\begin{center}
\normalsize \textbf{Marat Akhmet$^1$, Mehmet Onur Fen$^{2,}\symbolfootnote[1]{Corresponding Author. E-mail: monur.fen@gmail.com, Tel: +90 312 585 02 17}$, Ardak Kashkynbayev$^1$} \\
\vspace{0.2cm}
\textit{\textbf{$^1$Department of Mathematics, Middle East Technical University, 06800 Ankara, Turkey}}

\vspace{0.1cm}
\textit{\textbf{$^2$Basic Sciences Unit, TED University, 06420 Ankara, Turkey}}
\vspace{0.1cm}
\end{center}

\vspace{0.3cm}

\begin{center}
\textbf{Abstract}
\end{center}

\vspace{-0.2cm}

\noindent\ignorespaces

It is rigorously proved that the chaotic dynamics of the non-smooth system with relay function is persistent even if a chaotic perturbation is applied. We consider chaos in a modified Li-Yorke sense such that infinitely many almost periodic motions take place in its basis. It is demonstrated that the system under investigation possesses countable infinity of chaotic sets of solutions. Coupled Duffing oscillators are used to show the effectiveness of our technique, and simulations that support the theoretical results are represented. Moreover, a chaos control procedure based on the Ott-Grebogi-Yorke algorithm is proposed to stabilize the unstable almost periodic motions embedded in the chaotic attractor.

\vspace{0.2cm}
 
\noindent\ignorespaces \textbf{Keywords:} Persistence of chaos; Li-Yorke chaos; Almost periodic motions; Relay system; Chaos control; Duffing equation

\vspace{0.6cm}


\section{Introduction}

The word ``persistence'' is not popular for differential equations since it is not usual to say about persistence of periodic solutions against periodic perturbations as well as other forms of regular motions such as quasi-periodic and almost periodic solutions in a similar way. In the literature, these types of problems have been investigated as a part of synchronization \cite{Pik01}-\cite{Luo13}. Chaos is not an exceptional term in this row, and those results which we recognize as synchronization of chaos can be interpreted as a specific type of persistence of chaos \cite{Fujisaka83}-\cite{Gon04}. The specification is characterized through an asymptotic relation between solutions of coupled systems. In other words, persistence has not been considered by researchers explicitly, except under the mask of synchronization or entrainment \cite{Pik01}-\cite{Mettin93}. In this study, we consider synchronization in its ultimately generalized form, without any additional asymptotic conditions, considering only ingredients of Li-Yorke chaos \cite{Li75}. This is the main theoretical novelty of the present paper.

An extension of the original definition of Li and Yorke \cite{Li75} to dimensions greater than one was performed by Marotto \cite{Marotto78}. It was demonstrated in \cite{Marotto78} that a multidimensional continuously differentiable map possesses Li-Yorke chaos if it has a snap-back repeller. Moreover, generalizations of Li-Yorke chaos to mappings in Banach spaces and complete metric spaces can be found in \cite{Shi04,Shi05}. Besides, in the paper \cite{Akh14}, the Li-Yorke definition of chaos was modified in such a way that infinitely many periodic motions separated from the motions of the scrambled set are replaced with almost periodic ones. In the present paper, we will also consider the Li-Yorke chaos in this modified sense.

In paper \cite{Akh24}, we have considered unpredictability as a \textit{global} phenomenon in weather dynamics on the basis of connected Lorenz systems, and this extension of chaos was performed by means of perturbations of Lorenz systems by chaotic solutions of their counterparts. Our suggestion may be a key to explain why the weather unpredictability is observed everywhere. This is true also for unpredictability and lack of forecasting in economics \cite{Akh20a,Akh20b}. To complete the explanation of weather and economical unpredictability as global phenomena by the analysis of interconnected models, we need to argument persistence of chaos of a model against chaotic perturbations as solutions of another similar models. From this point of view, results of the present paper are very motivated. Another motivation of this study relies on the richness of a single chaotic model for motions, as a supply of infinitely many different periodic \cite{Li75,Dev90}, almost periodic \cite{Akh14} and even Poisson stable \cite{Akh21} motions. Of course, the diversity of motions is useless if one cannot control chaos  \cite{Ott90,Pyragas92}. In other words, chaos persistence means extension of chaos controllability.

In our former paper \cite{Fen17}, persistence of chaos was considered in coupled Lorenz systems by taking into account sensitivity and existence of infinitely many periodic motions embedded in the chaotic attractor. However, in the present paper, we consider chaos in the sense of Li-Yorke with countable infinity of almost periodic solutions in basis instead of periodic ones. In the present study, all results concerning the existence of almost periodic motions as well as Li-Yorke chaos are rigorously proved, and a more comprehensive theoretical discussion is performed compared to \cite{Fen17}. The demonstration of infinite number of Li-Yorke chaotic sets of solutions in the dynamics is another novelty of the present paper. Moreover, a numerical chaos control technique based on the Ott-Grebogi-Yorke (OGY) \cite{Ott90} algorithm is proposed for the stabilization of the unstable almost periodic motions. On the other hand, the paper \cite{Akh14} was concerned with the Li-Yorke chaotic dynamics of shunting inhibitory cellular neural networks with discontinuous external inputs. The concept of persistence of chaos was not considered in \cite{Akh14} at all. It was demonstrated in \cite{Akh14} that the chaotic structure of the discontinuity moments of the external inputs gives rise to the appearance of chaos, and chaos does not take place in the dynamics either in the case of regular discontinuity moments or in the absence of the discontinuous external inputs. On the contrary, in the present paper, a continuous chaotic perturbation is applied to a relay system which is already chaotic in the absence of perturbation, and it is proved that the chaotic structure is permanent in the dynamics regardless of the applied perturbation. As the source of chaotic perturbation we make use of solutions of another system of differential equations, but it is also possible to use any data which is known to be chaotic in the sense of Li-Yorke.

In the present study, we take into account the systems
\begin{eqnarray}\label{1}
x'=F(x,t)
\end{eqnarray}
and
\begin{eqnarray}\label{e1}
z'=Az+f(z,t)+ \nu (t,\zeta),
\end{eqnarray}
where $t\in \mathbb{R},$ the functions $F:\mathbb R^{m} \times \mathbb R\to \mathbb R^{m}$ and $f: \mathbb R^{n} \times  \mathbb R \to \mathbb R^n$ are continuous in all their arguments, $f(z,t)$ is almost periodic in $t$ uniformly for $z\in\mathbb R^n,$ $A \in \mathbb R^{n \times n}$ is a matrix whose eigenvalues have negative real parts, and 
\begin{equation} \label{relay}
\nu (t,\zeta)= \begin{cases}
m_0,& \textrm{if} \quad \zeta_{2i} < t \le \zeta_{2i+1}, \ i\in \mathbb{Z}, \\
m_1,& \textrm{if} \quad \zeta_{2i-1} < t \le \zeta_{2i}, \ i\in \mathbb{Z},
\end{cases}
\end{equation}
is a relay function in which $m_0,$ $m_1\in \mathbb{R}^n$ with $m_0\ne m_1.$ In (\ref{relay}), the sequence $\zeta=\left\{\zeta_{i}\right\},$ $i\in \mathbb{Z},$ of switching moments is defined through the equation 
\begin{eqnarray} \label{zeta_seq}
\zeta_{i} = \tau_i + \kappa_i,
\end{eqnarray} 
where $\left\{\tau_i^j\right\},$ $j\in\mathbb Z,$ is a family of equipotentially almost periodic sequences and the sequence $\left\{\kappa_i\right\},$ $\kappa_{0} \in \left[0, 1\right],$ is a solution of the logistic map 
\begin{eqnarray} \label{logistic}
\kappa_{i+1} = G_{\mu}(\kappa_i),
\end{eqnarray}
where $G_{\mu}(s) = \mu s(1-s)$ and $\mu$ is a parameter. Here, $\tau^j_i=\tau_{i+j}-\tau_{i}$ for each integers $i$ and $j.$

The presence of Li-Yorke chaos in the dynamics of (\ref{1}) is one of our main assumptions. We fix a value of $\mu$ between $3.84$ and $4$ such that the map (\ref{logistic}) is chaotic in the sense of Li-Yorke \cite{Li75}. For such a value of the parameter, the interval $\left[0, 1\right]$ is invariant under the iterations of (\ref{logistic}) \cite{Hale91}. An interpretation of the relay system (\ref{e1}) from the economic point of view can be found in \cite{Akh20a}. According to the results of papers \cite{Akh14,Akh1}, one can confirm that system (\ref{e1}) is Li-Yorke chaotic under certain conditions, which will be given in the next section.

We establish a unidirectional coupling between the systems (\ref{1}) and (\ref{e1}) to set up the following system,
\begin{eqnarray}\label{2}
y'=Ay+f(y,t)+\nu(t,\zeta)+h(x(t)),
\end{eqnarray}
where $x(t)$ is a solution of (\ref{1}), and $h:\mathbb R^m \to \mathbb R^n$ is a continuous function. Our purpose is to prove rigorously that the dynamics of system (\ref{2}) is Li-Yorke chaotic. In other words, we will show that the chaos of (\ref{e1}) is persistent even if it is perturbed with the solutions of system (\ref{1}).

Sufficient conditions on systems (\ref{1}), (\ref{e1}) and (\ref{2}) for the persistence of chaos, and the descriptions concerning almost periodicity and Li-Yorke chaos are provided in the next section.

\section{Preliminaries}

Throughout the paper, we will make use of the usual Euclidean norm for vectors and the norm induced by the Euclidean norm for matrices \cite{Horn92}. 

In our theoretical discussions, we will make use of the concept of Li-Yorke chaotic set of functions \cite{Akh14,Akh8,AkhF}. The description of the concept is as follows.

Suppose that $\Gamma \subset \mathbb R^p$ is a bounded set, and denote by
\begin{eqnarray} \label{collection}
\mathcal{H}=\left\{\psi(t)~|~ \psi: \mathbb R \to \Gamma ~\textrm{is}  ~\textrm{continuous}  \right\}
\end{eqnarray}
a collection of uniformly bounded functions.

A couple of functions $ \left( \psi(t), \overline{\psi}(t) \right) \in \mathcal{H} \times \mathcal{H}$ is called proximal if for arbitrary small $\epsilon>0$ and arbitrary large $E>0,$ there exists an interval $J$ with a length no less than $E$ such that $\left\|\psi(t)-\overline{\psi}(t)\right\| < \epsilon$ for each $t\in J$. On the other hand, we say that the couple $\left( \psi(t), \overline{\psi}(t) \right) \in \mathcal{H} \times \mathcal{H}$ is frequently $(\epsilon_0, \Delta)-$separated if there exist positive numbers $\epsilon_0, \Delta$ and infinitely many disjoint intervals with lengths no less than $\Delta$ such that $\left\|\psi(t)-\overline{\psi}(t)\right\| > \epsilon_0$ for each $t$ from these intervals. We call the couple $\left( \psi(t), \overline{\psi}(t) \right) \in \mathcal{H} \times \mathcal{H}$ a Li-Yorke pair if it is proximal and frequently $(\epsilon_0, \Delta)$-separated for some positive numbers $\epsilon_0$ and $\Delta.$ Moreover, a set $\mathcal{S} \subset \mathcal{H}$ is called a scrambled set if $\mathcal{S}$ does not contain any almost periodic function and each couple of different functions inside $\mathcal{S} \times \mathcal{S}$ is a Li-Yorke pair.

$\mathcal{H}$ is called a Li-Yorke chaotic set if: $(i)$ $\mathcal{H}$ admits a countably infinite subset of almost periodic functions; $(ii)$ there exists an uncountable scrambled set $ \mathcal{S}\subset \mathcal{H};$ $(iii)$ for any function $\psi(t)\in \mathcal{S}$ and any almost periodic function $\overline{\psi}(t)\in \mathcal{H},$ the pair $\left(\psi(t),\overline{\psi}(t)\right)$ is frequently $(\epsilon_0, \Delta)-$separated for some positive numbers $\epsilon_0$ and $\Delta.$

\begin{remark}
The criterion for the existence of a countably infinite subset of almost periodic functions in a Li-Yorke chaotic set can be replaced with the existence of a countably infinite subset of quasi-periodic or periodic functions.
\end{remark}

The presence of Li-Yorke chaos with a basis consisting of infinitely many almost periodic motions in the dynamics of system (\ref{1}) is one of our main assumptions. In other words, we suppose that system (\ref{1}) possesses a set $\mathscr{A}$ of uniformly bounded solutions which is chaotic in the Li-Yorke sense. In this case, there exists a compact region $\Lambda \subset \mathbb R^m$ such that the trajectories of all solutions that belong to $\mathscr{A}$ is inside $\Lambda.$ An example of such a system was provided in \cite{Akh14} with a theoretical discussion.

By means of the assumption on the matrix $A,$ one can confirm the existence of positive numbers $K$ and $\alpha$ such that $\|e^{A t}\|\le Ke^{-\alpha t},$ $t\ge 0.$ In the remaining parts of the paper, $\Theta$ will stand for the set of all sequences $\zeta=\left\{\zeta_i\right\}$, $i\in\mathbb Z$, generated by equation (\ref{zeta_seq}). Since the value of $\mu$ in (\ref{logistic}) is fixed between $3.84$ and $4$ so that the map is chaotic in the sense of Li-Yorke, the map (\ref{logistic}) possesses a periodic orbit with period $p$ for each natural number $p$ \cite{Li75}. We will denote by $\mathcal{P} \subset \Theta$ the countably infinite set of all sequences $\zeta=\left\{\zeta_i\right\}$, $i\in\mathbb Z,$ generated by (\ref{zeta_seq}) in which $\left\{\kappa_i\right\}$ is a periodic solution of (\ref{logistic}).

The following assumptions are required:
\begin{enumerate}
\item[\bf (C1)] There exists a positive number $M_f$ such that $\displaystyle \sup_{y\in \mathbb R^n, t\in \mathbb R} \left\|f(y,t)\right\|\le M_f$;
\item[\bf (C2)] There exists a positive number $L_f<\alpha/K$ such that $\left\|f(y_1,t)-f(y_2,t))\right\| \leq L_f\left\|y_1-y_2\right\|$ for all $y_1,$ $y_2 \in \mathbb R^n,$ $t\in \mathbb R$; 
\item[\bf (C3)] There exists a positive number $\underline{\zeta}$ such that $\zeta_{i+1}-\zeta_{i} \ge \underline{\zeta}$ for each $\zeta=\left\{\zeta_i\right\} \in \Theta$ and $i\in\mathbb Z$; 
\item[\bf (C4)]	There exists a positive number $L_1$ such that $\left\|h(x_1)-h(x_2)\right\| \le L_1\left\|x_1-x_2\right\|$ for all  $x_1, x_2 \in \Lambda$;
\item[\bf (C5)] There exists a positive number $L_2$ such that $\left\|h(x_1)-h(x_2)\right\| \ge L_2\left\|x_1-x_2\right\|$  for all $x_1, x_2 \in \Lambda$;
\item[\bf (C6)] There exists a positive number $M_{F}$ such that $\displaystyle \sup_{x\in \Lambda, t\in \mathbb R} \left\|F(x,t)\right\|\le M_{F}$. 
\end{enumerate}

Under the conditions $(C1)-(C3)$, one can confirm using the results of papers \cite{Akh14,Akh1} that the relay system (\ref{e1}) is Li-Yorke chaotic with infinitely many almost periodic motions in basis for the values of the parameter $\mu$ between $3.84$ and $4$. We refer the reader to \cite{AkhF} for further information about the dynamics of relay systems. 
Under the same conditions, for a given solution $x(t)\in\mathscr{A}$ of (\ref{1}) and a sequence $\zeta \in \Theta$, system (\ref{2}) possesses a unique solution $\phi_{x,\zeta}(t)$ which is bounded on the whole real axis \cite{Hale80}, and this solution satisfies the relation
\begin{equation}\label{e6}
\phi_{x,\zeta}(t)=\int_{-\infty}^t e^{A(t-s)}\left[f(\phi_{x,\zeta}(s),s)+\nu(s,\zeta)+h(x(s))\right]ds.
\end{equation}

For a fixed sequence $\zeta \in \Theta$ and a fixed solution $x(t)\in\mathscr{A},$ the bounded solution $\phi_{x,\zeta}(t)$ attracts all other solutions  of (\ref{2}) such that the inequality $\left\|\phi_{x,\zeta}(t) - y_{x,\zeta}(t)\right\| \leq K \left\|\phi_{x,\zeta}(t_0) - y_0\right\| e^{(KL_f-\alpha)(t-t_0)}$ is satisfied for $t\geq t_0,$ where $y_{x,\zeta}(t)$ is a solution of (\ref{2}) with $y_{x,\zeta}(t_0)=y_0$ for some $y_0\in\mathbb R^n$ and $t_0 \in \mathbb R.$

To provide a theoretical discussion for the persistence of chaos, for each sequence $\zeta\in\Theta$, let us introduce the set $\mathscr{B}_{\zeta}$ consisting of all bounded solutions $\phi_{x,\zeta}(t)$ of system (\ref{2}) in which $x(t)$ belongs to $\mathscr{A}$.

One can confirm  that $\displaystyle \sup_{t\in\mathbb R}\|y(t)\| \le M$ for all $y(t) \in \mathscr{B}_{\zeta}$ and $\eta\in\Theta$, where $ M=\displaystyle \frac{K}{\alpha}(\overline{m}+M_f+M_h)$, $\overline{m}=\max\left\lbrace \|m_0\|,\|m_1\| \right\rbrace$, and $M_h=\displaystyle \max_{ x\in \Lambda} \left\|h(x)\right\|$.

The next section is devoted to the almost periodic solutions of system (\ref{2}).

\section{Existence of almost periodic solutions}

Let $\{\sigma_{i}\}$, $i\in \mathbb Z$, be a sequence in $\mathbb{R}^{n}.$ An integer $p$ is an $\epsilon-$almost period of the sequence $\{\sigma_{i}\},$ if the inequality $\|\sigma_{i+p}-\sigma_{i}\|<\epsilon $ holds for all $i\in \mathbb{Z}.$ On the other hand, a set $\mathscr{D} \subset \mathbb{R} $ is said to be relatively dense if there exists a number $l>0$ such that $\big[r, r+l\big] \cap \mathscr{D} \neq \emptyset$ for all $r \in \mathbb{R}.$ Moreover, $\{\sigma_{i}\}$  is almost periodic, if for any $\epsilon>0,$ there exists a relatively dense set of its $\epsilon-$almost periods.

Let us denote $\xi_{i}^{j}=\xi_{i+j}-\xi_{i}$ for any integers $i$ and $j.$ We call the family of sequences $\{\xi_{i}^{j}\},$ ${j\in\mathbb{Z}},$  equipotentially almost periodic if for an arbitrary $\epsilon>0,$ there exists a relatively dense set of $\epsilon-$almost periods, common for all sequences $\{\xi_{i}^{j}\},$ $j\in\mathbb{Z}$ \cite{SP}.

A continuous function $\varphi:\mathbb R \to \mathbb R^n$ is said to be almost periodic if for any $\epsilon > 0$ there exists $l > 0$ such that for any interval with length $l$ there exists a number $\omega$ in this interval satisfying $\left\|\varphi(t+\omega)-\varphi(t) \right\| < \epsilon$ for all $ t \in \mathbb{R}$ \cite{Hale80}-\cite{Lev82}.

The following modified version of an assertion from \cite{ref2a} is needed for the proof of the main theorem of the present section.
\begin{lemma} \label{lemma2}
Suppose that $\varphi:\mathbb R\to\mathbb R^n$ is a continuous almost periodic function and $\left\{\xi_{i}^{j}\right\},$ ${j\in\mathbb{Z}},$ is a family of equipotentially almost periodic sequences. Then, for arbitrary $\eta>0 $ and $0<\theta<\eta,$ there exist relatively dense sets of real numbers $\Omega$ and even integers $Q$ such that 
\begin{enumerate}
\item[(i)] $\|\varphi(t+\omega)-\varphi(t)\|<\eta,$ $t\in \mathbb{R};$
\item[(ii)] $|\xi_{i}^{q}-\omega |<\theta,$ $i\in \mathbb{Z},$ $\omega \in \Omega,$ $q\in Q.$ 
\end{enumerate}
\end{lemma}

The existence of almost periodic solutions in system \eqref{2} is considered in the following theorem.

\begin{theorem} \label{almost}
Suppose that conditions $(C1)-(C4)$ are valid. If the sequence $\zeta=\left\{\zeta_i\right\}$, $i\in\mathbb Z$, belongs to $\mathcal{P}$ and $x(t)$ is an almost periodic solution of (\ref{1}), then the bounded solution $\phi_{x,\zeta}(t)$ is the unique almost periodic solution of (\ref{2}).
\end{theorem}

\noindent \textbf{Proof.}
Let us denote by $\mathscr{C}_0$ the set of all almost periodic functions $\psi:\mathbb R \to \mathbb R^n$ satisfying $\left\|\psi\right\|_{0}\leq M,$ where $\displaystyle \left\|\psi\right\|_0=\sup_{t\in\mathbb R}\left\|\psi(t)\right\|.$
Define the operator $ \Pi $ on $\mathscr{C}_0$ through the equation
$$\Pi \psi(t) = \int_{-\infty}^{t} e^{A(t-s)}\left(f(\psi(s),s)+v(s,\zeta)+h(x(s))\right) ds.$$

First of all, we will show that $\Pi(\mathscr{C}_0)\subseteq \mathscr{C}_0.$ Let $\psi$ be an element of $\mathscr{C}_0.$ One can easily verify that $\left\|\Pi \psi\right\|_0 \le M.$

In order to show that $\Pi\psi(t)$ is almost periodic, let us fix an arbitrary positive number $\epsilon$ and set $$H_0=K\left(\frac{1+L_f+L_1}{\alpha} + \frac{2\left\|m_0-m_1\right\|}{1-e^{-\alpha \underline{\zeta}}}\right).$$

Since $\Pi\psi$ is uniformly continuous, there exists a positive number $\eta$ satisfying $\eta < \displaystyle \frac{\underline{\zeta}}{5}$ and $\eta \leq \displaystyle \frac{\epsilon}{3H_0}$ such that $\left\|\Pi \psi(t_1) - \Pi \psi(t_2)\right\|<\displaystyle \frac{\epsilon}{3}$ whenever $\left|t_1-t_2\right|<4\eta.$
 
According to Lemma A.3 \cite{Akh14}, $\left\{\zeta_i^j\right\},$ $j\in\mathbb Z,$ is a family of equipotentially almost periodic sequences, since $\left\{\tau_i^j\right\},$ $j\in\mathbb Z,$ is a family of equipotentially almost periodic sequences and the sequence $\left\{\kappa_i\right\},$ $i\in\mathbb Z,$ is periodic. Let $\theta$ be a number with $0<\theta<\eta,$ and consider the numbers $\omega \in \Omega$ and $q \in Q$ as in Lemma \ref{lemma2} such that (i) $\left\|\psi(t+\omega)-\psi(t)\right\|<\eta,$ $t\in\mathbb R;$ (ii) $\left\|f(z,t+\omega) - f(z,t)\right\| <\eta,$ $z\in\mathbb R^n,$ $t\in\mathbb Z,$ (iii) $\left\|x(t+\omega) - x(t)\right\|<\eta,$ $t\in\mathbb R;$ (iv) $\left|\zeta_i^q-\omega\right|<\theta,$ $i\in\mathbb Z.$

For any $k\in\mathbb Z,$ it can be verified that if $s\in (\zeta_k+\theta, \zeta_{k+1}-\theta),$ then $\nu(s+\omega,\zeta)-\nu(s,\zeta)=0.$ 
Therefore, for each $t$ from the intervals $(\zeta_{i}+\eta, \zeta_{i+1}-\eta),$ $i\in\mathbb Z,$ one can confirm that 
\begin{eqnarray*} 
\left\|\Pi\psi(t+\omega) - \Pi \psi(t)\right\| < H_0 \eta \leq \frac{\epsilon}{3}.
\end{eqnarray*}  


Suppose that $t\in(\zeta_{\widetilde{i}}-\eta, \zeta_{\widetilde{i}}+\eta)$ for some $\widetilde{i}\in\mathbb Z.$ Because $\eta$ is sufficiently small such that $5\eta< \underline{\zeta},$ $t+3\eta$ belongs to the interval $(\zeta_{\widetilde{i}}+\eta, \zeta_{\widetilde{i}+1}-\eta)$ so that $$\left\|\Pi\psi(t+\omega+3\eta) - \Pi\psi(t+3\eta)\right\|<\displaystyle\frac{\epsilon}{3}.$$
 
Thus, we have that
\begin{eqnarray*}
	&& \|\Pi \psi(t+\omega)-\Pi \psi(t)\|  \leq  \|\Pi \psi(t+\omega)-\Pi \psi(t+\omega+3\eta)\| \\
	&& + \|\Pi \psi(t+\omega+3\eta)-\Pi \psi(t+3\eta)\|  + \|\Pi \psi(t+3\eta)-\Pi \psi(t)\|  \\
	&& < \epsilon.
\end{eqnarray*}
 
Therefore,  $\|\Pi \psi(t+\omega)-\Pi \psi(t)\| < \epsilon$ for all $t\in\mathbb R.$ Accordingly $\Pi \psi(t)$ is almost periodic and $\Pi(\mathscr{C}_0)\subseteq \mathscr{C}_0.$

Now, let $\psi_1$ and $\psi_2$ be elements of $\mathscr{C}_0.$ Then,
\begin{eqnarray*}
\left\|\Pi \psi_1(t) - \Pi \psi_2(t)\right\| \leq \displaystyle \int_{-\infty}^t K L_f  e^{-\alpha (t-s)} \left\|\psi_1(s)-\psi_2(s)\right\| ds \leq \displaystyle \frac{KL_f}{\alpha} \left\|\psi_1-\psi_2\right\|_0.
\end{eqnarray*}
Hence, $\displaystyle \left\|\Pi \psi_1-\Pi \psi_2\right\|_0 \leq \frac{KL_f}{\alpha} \left\|\psi_1-\psi_2\right\|_0.$ Since $\displaystyle \frac{KL_f}{\alpha}<1$, the operator $\Pi:\mathscr{C}_0\to \mathscr{C}_0$ is contractive according to condition $(C2).$ Consequently, the bounded solution $\phi_{x,\zeta}(t)$ is the unique almost periodic solution of system (\ref{2}). $\square$

We will consider the chaotic dynamics of system (\ref{2}) in the next section.

\section{Persistence of chaos}\label{main_lemmas}

The following lemmas, which are concerned with the proximality and frequent separation features of the bounded solutions of (\ref{2}), are needed for the proof of the main theorem of the present section.  

\begin{lemma} \label{proximality_lemma}
Suppose that conditions $(C1)-(C4)$ are valid. If a couple of functions  $\left( x(t), \widetilde{x}(t) \right) \in \mathscr{A}\times \mathscr{A}$ is proximal, then the same is true for the couple $ \left( \phi_{x,\zeta}(t),\phi_{\widetilde{x},\zeta}(t) \right) \in \mathscr{B}_{\zeta} \times \mathscr{B}_{\zeta}$ for any sequence $\zeta \in \Theta.$
\end{lemma}

\begin{lemma} \label{separation_lemma}
Suppose that conditions $(C1)-(C3)$, $(C5)$ and $(C6)$ are valid. If a couple of functions  $\left( x(t), \widetilde{x}(t) \right) \in \mathscr{A} \times \mathscr{A}$ is frequently $(\epsilon_{0},\Delta)-$separated for some positive numbers $\epsilon_{0}$ and $ \Delta $, then there exist positive numbers $\epsilon_{1}$ and $\overline{\Delta}$ such that the couple of functions $\left( \phi_{x,\zeta}(t), \phi_{\widetilde{x},\zeta}(t) \right) \in \mathscr{B}_{\zeta} \times \mathscr{B}_{\zeta}$ is frequently $(\epsilon_{1},\overline{\Delta})-$separated for any sequence $\zeta\in\Theta$.
\end{lemma}	

The proofs of Lemma \ref{proximality_lemma} and Lemma \ref{separation_lemma} are provided in the Appendix.
The main result of the present section is mentioned in the next theorem. 

\begin{theorem}\label{li-yorke_theorem}
Suppose that the conditions $(C1)-(C6)$ are valid. If $\zeta \in \mathcal{P}$, then $\mathscr{B}_{\zeta}$ is a Li-Yorke chaotic set.
\end{theorem}

\noindent \textbf{Proof.}
Since the set $\mathscr{A}$ is Li-Yorke chaotic, there exists a countably infinite set $\mathcal{AP}_1 \subset\mathscr{A}$ of almost periodic solutions. Fix an arbitrary sequence $\zeta \in \mathcal{P}$, and let us denote by $\mathcal{AP}^{\zeta}_2$ the subset of $\mathscr{B}_{\zeta}$ consisting of bounded solutions $\phi_{x,\zeta}(t)$ of (\ref{2}) such that $x(t) \in \mathcal{AP}_1$. According to Theorem \ref{almost}, the elements of $\mathcal{AP}^{\zeta}_2$ are the almost periodic solutions of system (\ref{2}). It can be verified using condition $(C5)$ that $\mathcal{AP}^{\zeta}_2$ is also countably infinite.

Now, suppose that $\mathcal{S}_{1} \subset \mathscr{A}$ is an uncountable scrambled set. Let us define the set of functions $\mathcal{S}^{\zeta}_2=\left\{\phi_{x,\zeta}(t): ~x(t) \in \mathcal{S}_1\right\}$. The set $\mathcal{S}^{\zeta}_2 \subset \mathscr{B}_{\zeta}$ is uncountable, and it does not contain any almost periodic solutions in accordance with condition $(C5)$. Since each couple of different functions inside $\mathcal{S}_1 \times \mathcal{S}_1$ is a Li-Yorke pair, Lemma \ref{proximality_lemma} and Lemma \ref{separation_lemma} together imply that the same is true for each couple of different functions inside $\mathcal{S}^{\zeta}_2 \times \mathcal{S}^{\zeta}_2.$ Hence, $\mathcal{S}^{\zeta}_2$ is a scrambled set. Besides, one can confirm using Lemma \ref{separation_lemma} one more time that each couple of functions inside $\mathcal{S}^{\zeta}_2 \times \mathcal{AP}^{\zeta}_2$ is frequently $\left(\epsilon_1,\overline{\Delta}\right)-$separated for some positive numbers $\epsilon_1$ and $ \overline{\Delta}$. Consequently, $\mathscr{B}_{\zeta}$ is a Li-Yorke chaotic set for each $\zeta \in \mathcal{P}$.
$\square$

\begin{remark}
In Theorem \ref{li-yorke_theorem}, we demonstrate that for each fixed sequence $\zeta \in \mathcal{P}$, system (\ref{2}) admits a Li-Yorke chaotic set $\mathscr{B}_{\zeta}$. It can be easily shown that $\mathscr{B}_{\zeta} \cap \mathscr{B}_{\eta} = \emptyset$ whenever $\zeta$ and $\eta$ are different sequences in $\mathcal{P}$. Therefore, there are countably infinite Li-Yorke chaotic sets in the dynamics of (\ref{2}).
\end{remark}

An illustrative example which supports the theoretical results is presented in the next section based on Duffing oscillators.

\section{An example}

Consider the forced Duffing equation
\begin{eqnarray} \label{Duffing1}
x''+1.5 x'+ 4x+0.02 x^3 = \cos t + \nu_1(t,\zeta),
\end{eqnarray}
where
\begin{equation} \label{ex_relay1}
\nu_1 (t,\zeta)= \begin{cases}
0.5,& \textrm{if} \quad \zeta_{2i} < t \le \zeta_{2i+1}, \ i\in \mathbb{Z}, \\
2.9,& \textrm{if} \quad \zeta_{2i-1} < t \le \zeta_{2i}, \ i\in \mathbb{Z},
\end{cases}
\end{equation}
is a relay function. The sequence $\zeta=\left\{\zeta_{i}\right\},$ $i\in \mathbb{Z},$ of switching moments is defined through the equation 
\begin{eqnarray}\label{examp_seq_zeta}
\zeta_{i} = 1.05 i + \kappa_i
\end{eqnarray} 
in which the sequence $\left\{\kappa_i\right\},$ $\kappa_0 \in [0,1]$, is a solution of the logistic map (\ref{logistic}). Clearly, $\zeta_{i+1}-\zeta_{i} \ge 0.05$ for each $i\in\mathbb Z.$

Using the variables $x_1=x$ and $x_2=x',$ one can write (\ref{Duffing1}) as a system in the following form,
\begin{eqnarray} \label{Duffing2}
\begin{array}{l}
x'_1=x_2, \\
x'_2=-4x_1 - 1.5 x_2 - 0.02x_1^3+ \cos t + \nu_1(t,\zeta).
\end{array}
\end{eqnarray}
The matrix of coefficients $\displaystyle \left( \begin{array}{ccc}
0 & 1 \\
-4 & -1.5  \end{array} \right)$ corresponding to the linear part of (\ref{Duffing2}) admits the eigenvalues $\displaystyle -\frac{3}{4} \pm i \frac{\sqrt{5}}{4}.$ The coefficient $-0.02$ of the nonlinear term in (\ref{Duffing2}) is sufficiently small in absolute value such that for each periodic orbit $\left\{\kappa_i\right\},$ $i\in\mathbb Z,$ of the map (\ref{logistic}), system (\ref{Duffing2}) possesses a unique quasi-periodic solution since the periods of the functions $\cos t$ and $\nu_1(t,\zeta)$ are incommensurable. In what follows, we will make use of the value $\mu=3.9$ so that system (\ref{Duffing2}) possesses Li-Yorke chaos with infinitely many quasi-periodic motions in basis \cite{Akh14,Akh1}.

\begin{figure}[ht]
\centering
\includegraphics[width=15cm]{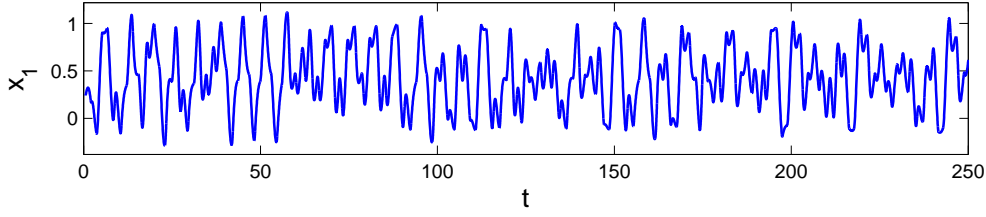}
\caption{Chaotic behavior in the $x_1-$coordinate of system (\ref{Duffing2}).}
\label{Fig1}
\end{figure}

In order to show the chaotic behavior of system (\ref{Duffing2}), in Figure \ref{Fig1} we depict the $x_1-$coordinate of the solution with the initial data $\zeta_0=0.56,$ $x_1(t_0)=0.24,$ $x_2(t_0)=0.17$ and $t_0=0.56.$ The simulation result seen in Figure \ref{Fig1} reveals the presence of chaos in the dynamics of (\ref{Duffing2}). One can numerically verify that the chaotic solutions of (\ref{Duffing2}) take place inside the compact region
\begin{eqnarray} \label{setlambda}
\Lambda= \left\{(x_1,x_2)\in\mathbb R^2: ~ -0.4 \leq x_1 \leq 1.25, ~ -1.3 \leq x_2 \leq 1.3 \right\}.
\end{eqnarray}

Next, we take into account the Duffing equation
\begin{eqnarray} \label{Duffing3}
z''+3.5 z'+ 2.5 z-0.01 z^3 = -1.5 \cos (\pi t) + \nu_2(t,\zeta).
\end{eqnarray}
In equation (\ref{Duffing3}), the relay function $\nu_2(t,\zeta)$ is defined as 
\begin{equation} \label{ex_relay2}
\nu_2 (t,\zeta)= \begin{cases}
1.7,& \textrm{if} \quad \zeta_{2i} < t \le \zeta_{2i+1}, \ i\in \mathbb{Z}, \\
-0.4,& \textrm{if} \quad \zeta_{2i-1} < t \le \zeta_{2i}, \ i\in \mathbb{Z},
\end{cases}
\end{equation}
in which the sequence $\zeta=\left\{\zeta_{i}\right\},$ $i\in \mathbb{Z},$ of switching moments is defined in the same way as in equation (\ref{examp_seq_zeta}).

Under the variables $z_1=z$ and $z_2=z',$ one can confirm that equation (\ref{Duffing3}) is equivalent to the system
\begin{eqnarray} \label{Duffing4}
\begin{array}{l}
z'_1=z_2, \\
z'_2=-2.5 z_1 - 3.5 z_2 + 0.01 z_1^3 -1.5 \cos (\pi t) + \nu_2(t,\zeta).
\end{array}
\end{eqnarray}
System (\ref{Duffing4}) is in the form of (\ref{e1}) with $A=\displaystyle \left( \begin{array}{ccc}
0 & 1 \\
-2.5 & -3.5 \end{array} \right)$
and
$f(z_1,z_2,t)=\displaystyle \left( \begin{array}{ccc}
0 \\
0.01z_1^3-0.5 \cos(\pi t) \end{array} \right).$
It can be verified that the eigenvalues of the matrix $A$ are $-1$ and $-5/2.$ Moreover, the inequality $\|e^{A t}\|\le Ke^{-\alpha t}$ holds for all $t\ge 0$ with $K=5.0695$ and $\alpha=1$. The coefficient $0.01$ of the nonlinear term in (\ref{Duffing4}) is sufficiently small such that according to the results of \cite{Akh1} system (\ref{Duffing4}) admits Li-Yorke chaos, but this time the basis of the chaos consists of infinitely many periodic motions since the periods of the functions $\cos (\pi t)$ and $\nu_2(t,\zeta)$ are commensurable for each periodic orbit $\left\{\kappa_i\right\}$, $i\in\mathbb Z$, of (\ref{logistic}).

\begin{figure}[ht]
\centering
\includegraphics[width=15cm]{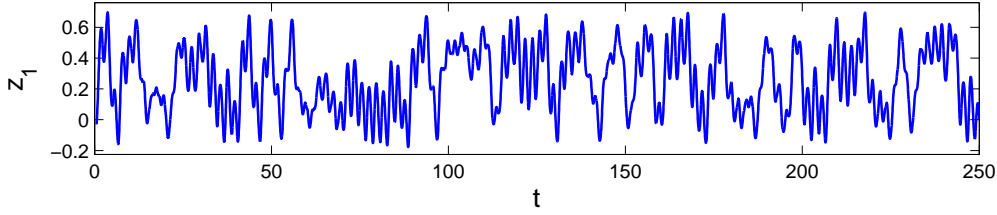}
\caption{The graph of the $z_1-$coordinate of system (\ref{Duffing4}). One can observe in the figure that system (\ref{Duffing4}) admits Li-Yorke chaos.}
\label{Fig2}
\end{figure}

Figure \ref{Fig2} represents the $z_1-$coordinate of the solution of (\ref{Duffing4}) corresponding to the initial data $\zeta_0=0.56,$ $z_1(t_0)=-0.03,$ $z_2(t_0)=0.32,$ where $t_0=0.56.$ One can observe in Figure \ref{Fig2} that Li--Yorke chaos takes place in the dynamics of system (\ref{Duffing4}).

Now, to demonstrate the persistence of chaos, we establish a unidirectional coupling between the systems (\ref{Duffing2}) and (\ref{Duffing4}) to set up the system
\begin{eqnarray} \label{Duffing5}
\begin{array}{l}
y'_1=y_2 + 0.7 x_1^2 -1.2 x_1, \\
y'_2=-2.5 y_1 - 3.5 y_2 + 0.01 y_1^3 -1.5 \cos (\pi t) + \nu_2(t,\zeta) + \arctan(x_2).
\end{array}
\end{eqnarray}

System (\ref{Duffing5}) is in the form of (\ref{2}) with $h(x_1,x_2)=\displaystyle \left( \begin{array}{ccc}
0.7 x_1^2 -1.2 x_1 \\
\arctan(x_2) \end{array} \right).$
It can be numerically verified that the bounded solutions of (\ref{Duffing5}) lie inside the compact region
$$
\mathcal{R}=\left\{(y_1,y_2) \in \mathbb R^2: ~ -0.9 \leq y_1 \leq 0.8, ~ -0.65 \leq y_2 \leq 1.2 \right\}.
$$
Therefore, the conditions $(C1)$ and $(C2)$ are valid for (\ref{Duffing5}) with $M_f=1.50729$ and $L_f=0.0243$. Moreover, the function $h(x_1,x_2)$ clearly satisfies the conditions $(C4)$ and $(C5)$ on the compact region $\Lambda$ defined by (\ref{setlambda}). The chaos of system (\ref{Duffing4}) is persistent under the applied perturbation such that system (\ref{Duffing5}) possesses Li-Yorke chaos with infinitely many almost periodic motions in basis according to Theorem \ref{li-yorke_theorem}. 

\begin{figure}[htp]
\centering
\includegraphics[width=15cm]{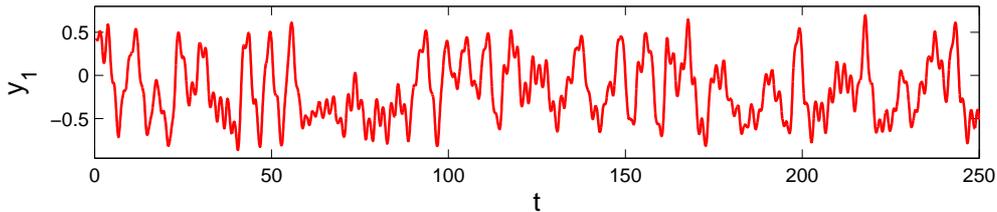}
\caption{The time series of the $y_1-$coordinate of system (\ref{Duffing5}). The figure reveals the persistence of chaos.}
\label{Fig3}
\end{figure}

\begin{figure}[htp]
\centering
\includegraphics[width=8.5cm]{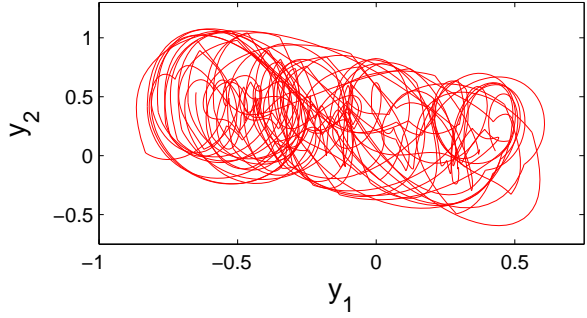}
\caption{The chaotic trajectory of system (\ref{Duffing5}). The simulation supports the result of Theorem \ref{li-yorke_theorem} such that the system remains to be chaotic even if the system is perturbed with the solutions of (\ref{Duffing2}).}
\label{Fig4}
\end{figure}

Using the solution $(x_1(t),x_2(t))$ of (\ref{Duffing2}) whose first coordinate is represented in Figure \ref{Fig1} in the perturbation, we depict in Figure \ref{Fig3} the $y_1-$coordinate of the solution of (\ref{Duffing5}) corresponding to the initial data $\zeta_0=0.56,$ $y_1(t_0)=0.43,$ $y_2(t_0)=0.04,$ where $t_0=0.56.$ Furthermore, Figure \ref{Fig4} shows the trajectory of the same solution on the $y_1-y_2$ plane. Both of Figures \ref{Fig3} and $\ref{Fig4}$ support the result of Theorem \ref{li-yorke_theorem} such that Li-Yorke chaos is permanent in the dynamics of (\ref{Duffing4}) even if the perturbation $h(x_1,x_2)$ is applied. 

\section{Control of chaos}

In this part of the paper, we will present a numerical technique to control the chaos of system (\ref{Duffing5}). For that purpose, we will make use of the OGY control method \cite{Ott90} applied to the logistic map (\ref{logistic}), which is the main source of chaos in the unidirectionally coupled system (\ref{Duffing2})+(\ref{Duffing5}).

Let us briefly explain the OGY algorithm for the logistic map (\ref{logistic}) \cite{Ott90,Sch99}. Denote by $\kappa^{(j)},$ $j=1,2,\ldots ,p,$ the target $p-$periodic orbit of (\ref{logistic}) with $\mu=3.9$ to be stabilized. Suppose that the parameter $\mu$ in (\ref{logistic}) is allowed to vary in the range $[3.9-\varepsilon, 3.9+\varepsilon]$, where $\varepsilon$ is a given small positive number. 
In the OGY control method \cite{Sch99}, after the control mechanism is switched on, we consider (\ref{logistic}) with the parameter value $\mu=\overline \mu_i,$ where
\begin{eqnarray} \label{control}
\overline \mu_i=3.9 \left(1+\frac{(2\kappa^{(j)}-1)(\overline{\kappa}_{i}-\kappa^{(j)})}{\kappa^{(j)}(1-\kappa^{(j)})} \right),
\end{eqnarray}
provided that the number on the right-hand side of the formula (\ref{control}) belongs to the interval $[3.9-\varepsilon, 3.9+\varepsilon].$ Here, the sequence $\left\{\overline{\kappa}_i\right\}$, $i\geq 0$, satisfying $\overline{\kappa}_0\in[0,1]$ is an arbitrary solution of the map
\begin{eqnarray} \label{control_kappa}
\overline{\kappa}_{i+1} = G_{\overline{\mu}_i}(\overline{\kappa}_i).
\end{eqnarray}
Formula (\ref{control}) is valid only if the trajectory $\left\{\overline{\kappa}_i\right\}$ is sufficiently close to the target periodic orbit $\kappa^{(j)},$ $j=1,2,\ldots,p$. Otherwise, we take $\overline \mu_{i}=3.9$ in (\ref{control_kappa}) so that the system evolves at its original parameter value, and wait until the trajectory $\left\{\overline{\kappa}_i\right\}$ enters in a sufficiently small neighborhood of the periodic orbit such that the inequality 
$-\varepsilon \le 3.9 \displaystyle\frac{(2\kappa^{(j)}-1)(\overline{\kappa}_{i}-\kappa^{(j)})}{\kappa^{(j)}(1-\kappa^{(j)})} \le \varepsilon$ holds. If this is the case, the control of chaos is not achieved immediately after switching on the control mechanism. Instead, there is a transition time before the desired periodic orbit is stabilized. The transition time increases if the number $\varepsilon$ decreases \cite{Gon04}.

Consider the sequence $\overline{\zeta}=\left\{\overline{\zeta}_i\right\}$, $i \geq 0$, generated through the equation $\overline{\zeta}_i = 1.05 i + \overline{\kappa}_i$, where $\left\{\overline\kappa_i\right\}$, $\overline{\kappa}_0 \in [0,1]$, is a solution of (\ref{control_kappa}).
To control the chaos of (\ref{Duffing5}), we replace the sequence $\zeta=\left\{\zeta_i\right\}$ in the coupled system (\ref{Duffing2})+(\ref{Duffing5}) with $\overline{\zeta}$, and consider the following control system conjugate to (\ref{Duffing2})+(\ref{Duffing5}):  
\begin{eqnarray} \label{Duffing6}
\begin{array}{l}
w'_1=w_2, \\
w'_2=-4w_1 - 1.5 w_2 - 0.02w_1^3+ \cos t + \nu_1\Big(t,\overline{\zeta}\Big), \\
w'_3=w_4 + 0.7 w_1^2 -1.2 w_1, \\
w'_4=-2.5 w_3 - 3.5 w_4 + 0.01 w_3^3 -1.5 \cos (\pi t) + \nu_2\Big(t,\overline{\zeta}\Big) + \arctan(w_2).
\end{array}
\end{eqnarray}

\begin{figure}[htp]
\centering
\includegraphics[width=15cm]{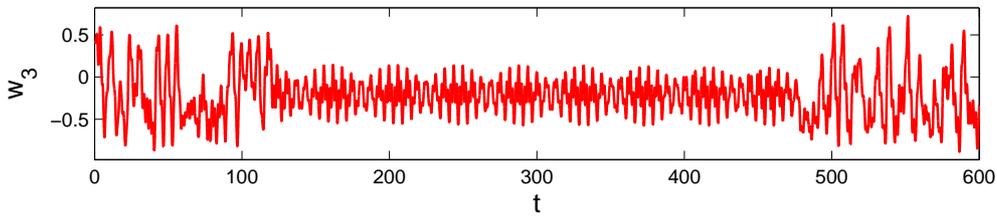}
\caption{Control of the chaos of (\ref{Duffing5}) by means of the OGY method applied to the map (\ref{logistic}) around its fixed point $2.9/3.9.$ The value $\varepsilon=0.08$ is used, and the control is switched on at $t=\zeta_{50}$ and switched off at $t=\zeta_{400}.$}
\label{Fig5}
\end{figure}

\begin{figure}[htp]
\centering
\includegraphics[width=15cm]{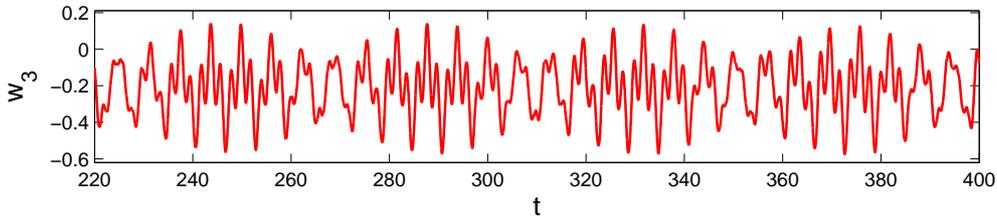}
\caption{The stabilized quasi-periodic solution of (\ref{Duffing5}).}
\label{Fig6}
\end{figure}

Figure \ref{Fig5} represents the time series of the $w_3-$coordinate of the control system (\ref{Duffing6}) corresponding to the initial data $\zeta_0=0.56$, $w_1(t_0)= 0.24$, $w_2(t_0)= 0.17$, $w_3(t_0)= 0.43$, $w_4(t_0)=0.04,$ where $t_0=0.56$. The OGY algorithm is applied around the fixed point $2.9/3.9$ of the logistic map (\ref{logistic}) by setting $\kappa^{(j)}\equiv 2.9/3.9$ in equation (\ref{control}). The control is switched on at $t=\zeta_{50}$ and switched off at $t=\zeta_{400},$ i.e., we take $\overline{\mu}_i=3.9$ for $0 \leq i <50$ and $i \geq 400$ in (\ref{control_kappa}). Moreover, the value $\varepsilon=0.08$ is used in the simulation. One can observe in Figure \ref{Fig5} that one of the quasi-periodic solutions embedded in the chaotic attractor of (\ref{Duffing5}) is stabilized. A transient time occurs after the control is switched on such that the stabilization becomes dominant approximately at $t=124$ and prolongs approximately till $t=477$ after which the chaotic behavior develops again. Moreover, the stabilized quasi-periodic solution of (\ref{Duffing5}) is represented in Figure \ref{Fig6}. Figures \ref{Fig5} and \ref{Fig6} manifest that the proposed numerical technique, which is based on the OGY algorithm, is appropriate to control the chaos of system (\ref{Duffing5}).

\section{Conclusions}

In this paper, persistence of chaos in a single model against chaotic perturbations arising exogenously has been considered. The problem seems difficult to be solved if one does not utilize the known methods of synchronization, since there is a possibility of compensation of chaos  by chaos. This is the reason why we have considered the problem more delicately taking into account the structure of the chaos under perturbations. The results can be loosely formulated as the coexistence of infinitely many Li-Yorke chaotic sets in the dynamics of coupled systems. The presence of quasi-periodic motions embedded in the chaotic attractor of systems of the form (\ref{2}) is confirmed by a numerical chaos control technique based on the OGY algorithm \cite{Ott90, Sch99}. Our results are applicable to a wide variety of systems including mechanical, electrical and economic models without any restriction in the dimension.

Suppression of chaos is considered in the literature by means of regular perturbations \cite{Gon04,Lima90,Braiman91}, while in the present study we discuss the persistence of chaos under chaotic perturbations. It is clear that persistence of chaos can be regarded as somehow opposite to the suppression of chaos, and they are complements in the theory. Likewise the importance of chaos suppression in applied sciences and technology \cite{Gon04}, the concept of persistence of chaos is also of great importance in such fields.

\section*{Appendix}

\textbf{Proof of Lemma \ref{proximality_lemma}}
 
Let us take a positive number $ \gamma $ satisfying $\gamma \le \displaystyle \left[ \frac{KL_1}{\alpha-KL_f}+\frac{2K(M_f+M_h)}{\alpha}\right]^{-1}$. Fix an arbitrary small positive number $ \epsilon $ and an arbitrary large real number $E$ satisfying $\displaystyle E \geq \frac{2}{\alpha-KL_f} \ln \left(\frac{1}{\gamma \epsilon}\right)$. Because the couple $\left( x(t), \widetilde{x}(t) \right) \in \mathscr{A}\times \mathscr{A}$ is proximal, there exist real numbers $R$ and $E_0 \ge E$ such that $\left\| x(t)-\widetilde{x}(t) \right\| < \gamma \epsilon$ for each $t \in [R,R+E_0]$. 

Fix an arbitrary sequence $\zeta\in\Theta.$ Using the relations
\begin{equation*} 
\phi_{x,\zeta}(t)=\int_{-\infty}^t e^{A(t-s)}\left[f(\phi_{x,\zeta}(s),s)+\nu(s,\zeta)+h(x(s))\right]ds
\end{equation*}
and
\begin{equation*} 
\phi_{\widetilde{x},\zeta}(t)=\int_{-\infty}^t e^{A(t-s)}\left[f(\phi_{\widetilde{x},\zeta}(s),s)+\nu(s,\zeta)+h(\widetilde{x}(s))\right]ds,
\end{equation*}
one can verify for $t \in [R, R+E_0]$ that
\begin{eqnarray} \label{proxproof_ineq}
\begin{array}{l}
\|\phi_{x,\zeta}(t)-\phi_{\widetilde{x},\zeta}(t)\|  \le \displaystyle \frac{2K(M_f+M_h)}{\alpha}e^{-\alpha (t-R)} + \frac{KL_1\gamma \epsilon}{\alpha} \left(1-e^{-\alpha(t-R)}\right) \\
 + KL_f \displaystyle \int_{R}^{t}e^{-\alpha (t-s)}\|\phi_{x,\zeta}(s)-\phi_{\widetilde{x},\zeta}(s)\|ds.
\end{array}
\end{eqnarray}

Let us introduce the function $w(t)=e^{\alpha t}\|\phi_{x,\zeta}(t)-\phi_{\widetilde{x},\zeta}(t)\|$.
Inequality (\ref{proxproof_ineq}) implies that
\begin{equation*}
w(t) \leq \frac{2K(M_f+M_h) - KL_1\gamma \epsilon}{\alpha}e^{\alpha R} + \frac{KL_1\gamma\epsilon}{\alpha}e^{\alpha t} + KL_f \displaystyle \int_{R}^{t}w(s)ds.
\end{equation*}
Applying Lemma $2.2$ \cite{bar} to the last inequality we obtain that
\begin{eqnarray*}
  w(t)  \leq \displaystyle \frac{KL_1 \gamma \epsilon}{\alpha-KL_f} e^{\alpha t} \left(1-e^{(KL_f-\alpha)(t-R)}\right)   
  + \frac{2K(M_f+M_h)}{\alpha} e^{KL_f t} e^{-(KL_f-\alpha)R}. 
\end{eqnarray*}
Therefore, we have for $t\in [R,R+E_0]$ that
$$
\|\phi_{x,\zeta}(t)-\phi_{\widetilde{x},\zeta}(t)\|  < \frac{KL_1\gamma\epsilon}{\alpha} + \frac{2K(M_f+M_h)}{\alpha} e^{(KL_f-\alpha)(t-R)}.
$$
Since the number $E$ is sufficiently large such that $\displaystyle E \geq \frac{2}{\alpha-KL_f} \ln \left(\frac{1}{\gamma \epsilon}\right)$, if $t\in [R+E/2,R+E_0]$, then $e^{(KL_f-\alpha)(t-R)} \leq \gamma\epsilon$. Thus,
$$
\|\phi_{x,\zeta}(t)-\phi_{\widetilde{x},\zeta}(t)\| < \displaystyle \left[ \frac{KL_1}{\alpha-KL_f}+\frac{2K(M_f+M_h)}{\alpha}\right] \gamma \epsilon \leq \epsilon
$$
for $t\in[R+E/2,R+E_0]$. It is worth noting that the interval $[R+E/2,R+E_0]$ has a length no less than $E/2$. Consequently, the couple $\left( \phi_{x,\zeta}(t), \phi_{\tilde{x},\zeta}(t) \right) \in \mathscr{B}_{\zeta} \times \mathscr{B}_{\zeta}$ is proximal for any sequence $\zeta\in\Theta$. $ \square $

\textbf{Proof of Lemma \ref{separation_lemma}}

Since the couple $\left( x(t), \widetilde{x}(t) \right) \in \mathscr{A} \times \mathscr{A}$ is frequently $(\epsilon_{0},\Delta)-$separated, there exist infinitely many disjoint intervals $I_k,$ $k\in\mathbb N,$ with lengths no less than $\Delta$ such that $\left\| x(t)-\widetilde{x}(t) \right\| > \epsilon_0$ for each $t$ from these intervals. According to condition $(C6)$ the set of functions $\mathscr{A}$ is an equicontinuous family on $\mathbb R$. Therefore, using the uniform continuity of the function $g:\Lambda \times \Lambda \to \mathbb R^n$ defined as $g(x_1,x_2)=h(x_1)-h(x_2),$ one can confirm that the family
$$ 
\mathcal{U}=\left\{h(x(t))-h(\widetilde{x}(t)):~  x(t)\in \mathscr{A},~ \widetilde{x}(t) \in \mathscr{A} \right\} 
$$ 
is also equicontinuous on $\mathbb R.$ Suppose that $h(x) = (h_1(x), h_2(x), \ldots, h_n(x)),$ where each $h_j,$ $j=1,2,\ldots,n,$ is a real valued function. In accordance with the equicontinuity of the family $\mathcal{U}$, there exists a positive number $\tau<\Delta,$ which does not depend on $x(t)$ and $\widetilde{x}(t)$, such that for any $t_1,t_2\in \mathbb R$ with $\left|t_1-t_2\right|<\tau$ we have 
\begin{eqnarray}\label{proof_eqn_1}
\left| \left(h_j\left(x(t_1)\right) - h_j\left(\widetilde{x}(t_1)\right) \right) - \left(h_j\left(x(t_2)\right) - h_j\left(\widetilde{x}(t_2)\right) \right) \right|<\frac{L_2\epsilon_0}{2\sqrt{n}}
\end{eqnarray}
for each $j=1,2,\ldots, n$.

Fix an arbitrary natural number $k$. Let us denote by $s_k$ be the midpoint of the interval $I_k$, and set $\eta_k=s_k-\tau/2.$ One can find an integer $j_k$ with $1 \leq j_k \leq n$ such that 
\begin{eqnarray} \label{proof_eqn_2}
\displaystyle \left|h_{j_k}(x(s_k))-h_{j_k}(\widetilde{x}(s_k))\right| \geq \frac{L_2}{\sqrt{n}} \left\|x(s_k)-\widetilde{x}(s_k)\right\| > \frac{L_2\epsilon_0}{\sqrt{n}}. 
\end{eqnarray}
Using (\ref{proof_eqn_1}) one can obtain for $t \in \left[\eta_k, \eta_k+\tau\right]$ that
\begin{eqnarray*}
\left|h_{j_k}\left(x(s_k)\right) - h_{j_k}\left(\widetilde{x}(s_k)\right) \right| - \left|h_{j_k}\left(x(t)\right) - h_{j_k}\left(\widetilde{x}(t)\right) \right| 
<\frac{L_2\epsilon_0}{2\sqrt{n}}.
\end{eqnarray*}
Moreover, inequality (\ref{proof_eqn_2}) yields
\begin{eqnarray*} 
 \left|h_{j_k}\left(x(t)\right) - h_{j_k}\left(\widetilde{x}(t)\right) \right| 
 > \left|h_{j_k}\left(x(s_k)\right) - h_{j_k}\left(\widetilde{x}(s_k)\right) \right|  - \frac{L_2\epsilon_0}{2\sqrt{n}}
 > \frac{L_2\epsilon_0}{2\sqrt{n}}
\end{eqnarray*}
for all $t \in \left[\eta_k, \eta_k+\tau\right].$ Since there exist numbers $r_1, r_2, \ldots, r_n \in [\eta_k, \eta_k + \tau]$ satisfying
\begin{eqnarray*}
\left\|\displaystyle \int_{\eta_k}^{\eta_k + \tau}  \left[ h(x(s)) - h(\widetilde{x}(s)) \right]  ds \right\| = \tau \left(\displaystyle \sum_{k=1}^n \left[h_k(x(r_k)) - h_k(\widetilde{x}(r_k))\right]  \right)^{1/2},
\end{eqnarray*}
we have that
\begin{eqnarray*}
&\left\|\displaystyle \int_{\eta_k}^{\eta_k + \tau}  \left[ h(x(s)) - h(\widetilde{x}(s)) \right]  ds \right\| & \geq  \tau \left|h_{j_k}(x(r_{j_k})) - h_{j_k}(\widetilde{x}(r_{j_k}))\right| \\
&& > \displaystyle  \frac{\tau L_2\epsilon_0}{2\sqrt{n}}.
\end{eqnarray*}

Let $\zeta\in\Theta$ be an arbitrary sequence. Making use of the relations
\begin{eqnarray*}
\phi_{x,\zeta}(t) = \phi_{x,\zeta}(\eta_k) + \displaystyle \int_{\eta_k}^t \left[A \phi_{x,\zeta}(s) + f(\phi_{x,\zeta}(s),s) + \nu(s,\zeta) + h(x(s))  \right] ds
\end{eqnarray*}
and
\begin{eqnarray*}
\phi_{\widetilde{x},\zeta}(t) = \phi_{\widetilde{x},\zeta}(\eta_k) + \displaystyle \int_{\eta_k}^t \left[A \phi_{\widetilde{x},\zeta}(s) + f(\phi_{\widetilde{x},\zeta}(s),s) + \nu(s,\zeta) + h(\widetilde{x}(s))  \right] ds,
\end{eqnarray*}
it can be deduced that
\begin{eqnarray*}
& \left\|\phi_{x,\zeta} (\eta_k + \tau)  - \phi_{\widetilde{x},\zeta} (\eta_k+\tau) \right\| &\geq \left\|  \displaystyle \int_{\eta_k}^{\eta_k+\tau}  \left[h(x(s)) - h(\widetilde{x}(s))\right] ds \right\| \\
&& - \left\|\phi_{x,\zeta} (\eta_k )  - \phi_{\widetilde{x},\zeta} (\eta_k) \right\| \\
&& - \left(\left\|A\right\|+L_f\right)\displaystyle \int_{\eta_k}^{\eta_k+\tau} \left\|\phi_{x,\zeta} (s)  - \phi_{\widetilde{x},\zeta} (s) \right\| ds.
\end{eqnarray*}
Hence,
$$
\displaystyle \max_{t\in[\eta_k,\eta_k+\tau]}  \left\| \phi_{x,\zeta}(t) - \phi_{\widetilde{x},\zeta}(t) \right\| > \displaystyle \frac{\tau L_2 \epsilon_0}{2[2+\tau(\left\|A\right\|+L_f)]\sqrt{n}}.
$$
Suppose that $\displaystyle \max_{t\in[\eta_k,\eta_k+\tau]}  \left\| \phi_{x,\zeta}(t) - \phi_{\widetilde{x},\zeta}(t) \right\|$ takes its maximum at the point $\lambda_k$ on the interval $[\eta_k,\eta_k+\tau]$. 

Define the positive numbers
$$
\displaystyle \overline{\Delta} = \min\left\{ \frac{\tau}{2}, \frac{\tau L_2 \epsilon_0}{4[M(\left\|A\right\|+L_f) + M_h] [2+\tau(\left\|A\right\|+L_f)] \sqrt{n}}\right\},
$$
and
$$\epsilon_1=\displaystyle \frac{\tau L_2 \epsilon_0}{4[2+\tau(\left\|A\right\|+L_f)]\sqrt{n}}.$$
Moreover, let $\delta_k=\lambda_k$ if $\lambda_k \in [\eta_k,\eta_k+\tau/2]$ and $\delta_k=\lambda_k-\overline{\Delta}$ if $\lambda_k \in (\eta_k+\tau/2,\eta_k+\tau]$.

The relation
\begin{eqnarray*}
& \phi_{x,\zeta}(t) - \phi_{\widetilde{x},\zeta}(t) &= \phi_{x,\zeta}(\lambda_k) - \phi_{\widetilde{x},\zeta}(\lambda_k) + \displaystyle \int_{\lambda_k}^t A \left[\phi_{x,\zeta}(s) - \phi_{\widetilde{x},\zeta}(s) \right] ds \\
&& + \displaystyle \int_{\lambda_k}^t \left[ f(\phi_{x,\zeta}(s),s) - f(\phi_{\widetilde{x},\zeta}(s),s) \right] ds \\
&& +  \displaystyle \int_{\lambda_k}^t \left[h(x(s)) - h(\widetilde{x}(s))\right] ds
\end{eqnarray*}
implies that
$
\left\|\phi_{x,\zeta}(t) - \phi_{\widetilde{x},\zeta}(t)\right\| > \epsilon_1
$
for $t\in\Big[\delta_k,\delta_k+\overline{\Delta}\Big]$.
 
Clearly, the intervals $\Big[\delta_k,\delta_k+\overline{\Delta}\Big]$, $k\in\mathbb N$, are disjoint. Consequently, the couple of functions $\left( \phi_{x,\zeta}(t), \phi_{\widetilde{x},\zeta}(t) \right) \in \mathscr{B}_{\zeta} \times \mathscr{B}_{\zeta}$ is frequently $(\epsilon_{1},\overline{\Delta})-$separated for any sequence $\zeta\in\Theta$. $\square$


\end{document}